\documentclass[11pt,preprint]{aastex}
\slugcomment{Accepted by {\it The Astrophysical Journal}}
\def\farcm{\hbox{$.\mkern-4mu^\prime$}}
\def\farcs{\hbox{$.\mkern-4mu^{\prime\prime}$}}

\def\la{\mathrel{\hbox{\rlap{\hbox{\lower4pt\hbox{$\sim$}}}\hbox{$<$}}}}
\def\ga{\mathrel{\hbox{\rlap{\hbox{\lower4pt\hbox{$\sim$}}}\hbox{$>$}}}}

\shortauthors{Park}
\shorttitle{N49}

\begin{document}
\title{An X-Ray Study of Supernova Remnant N49 and Soft Gamma-Ray Repeater 0526--66 in 
the Large Magellanic Cloud}

\author{Sangwook Park\altaffilmark{1}, John P. Hughes\altaffilmark{2},
Patrick O. Slane\altaffilmark{3}, David N. Burrows\altaffilmark{4},
Jae-Joon Lee\altaffilmark{5}, and Koji Mori\altaffilmark{6}} 

\altaffiltext{1}{Box 19059, Department of Physics, University of Texas at Arlington,
Arlington, TX 76019; s.park@uta.edu}
\altaffiltext{2}{Department of Physics and Astronomy, Rutgers University,
136 Frelinghuysen Road, Piscataway, NJ 08854-8019}
\altaffiltext{3}{Harvard-Smithsonian Center for Astrophysics, 60 Garden Street,
Cambridge, MA 02138}
\altaffiltext{4}{Department of Astronomy and Astrophysics, Pennsylvania State
University, 525 Davey Laboratory, University Park, PA 16802}
\altaffiltext{5}{Korea Astronomy and Space Science Institute, Daejeon, 305-348, Korea}
\altaffiltext{6}{Department of Applied Physics, University of Miyazaki, 1-1 Gakuen 
Kibana-dai Nishi, Miyazaki, 889-2192, Japan} 


\begin{abstract}

We report on the results from our deep {\it Chandra} observation (120 ks) of the supernova 
remnant (SNR) N49 and soft $\gamma$-ray repeater (SGR) 0526--66 in the Large Magellanic 
Cloud. We firmly establish the detection of an ejecta ``bullet'' beyond the southwestern 
boundary of N49. The X-ray spectrum of the bullet is distinguished from that of the main SNR 
shell, showing significantly enhanced Si and S abundances. We also detect an ejecta feature 
in the eastern shell, which shows metal overabundances similar to those of the bullet. If N49 
was produced by a core-collapse explosion of a massive star, the detected Si-rich ejecta may 
represent explosive O-burning or incomplete Si-burning products from deep interior of the 
supernova. On the other hand, the observed Si/S abundance ratio in the ejecta may favor Type 
Ia origin for N49. We refine the Sedov age of N49, $\tau_{\rm Sed}~\sim$~4800~yr, with the 
explosion energy $E_0~\sim$~1.8~$\times$~10$^{51}$~erg. Our blackbody (BB) + power law (PL) 
model for the quiescent X-ray emission from SGR 0526--66 indicates that the PL photon index 
($\Gamma~\sim$~2.5) is identical to that of PSR 1E1048.1--5937, the well-known candidate 
transition object between anomalous X-ray pulsars and SGRs.  Alternatively, the two-component 
BB model implies X-ray emission from a small ($R~\sim$~1~km) hot spot(s) ($kT~\sim$~1~keV) 
in addition to emission from the neutron star's cooler surface ($R~\sim$~10~km, 
$kT~\sim$~0.4~keV). There is a considerable discrepancy in the estimated column toward 
0526--66 between BB+PL and BB+BB model fits. Discriminating these spectral models would 
be crucial to test the long-debated physical association between N49 and 0526--66. 

\end{abstract}

\keywords {ISM: supernova remnants --- ISM: individual objects: N49 --- Stars: magnetars --- 
Stars: individual: SGR 0526--66 --- X-rays: ISM --- X-rays: stars}

\section {\label {sec:intro} INTRODUCTION}

The supernova remnant (SNR) N49 is the third brightest X-ray SNR in the Large Magellanic Cloud 
(LMC). The blast wave is interacting with dense clumpy interstellar clouds on the SNR's eastern
side \citep{vanc92,bana97}, producing bright emission in optical, infrared, ultraviolet, and 
X-ray bands \citep{park03a,sank04,will06,rako07,bili07,otsu10}, which revealed the shock structures 
in N49 down to sub-pc scales over wide ranges of the velocity ($v$ $\sim$ 10$^{2-3}$ km s$^{-1}$), 
temperature ($T$ $\sim$ 10$^{2-7}$ K), and the interacting density ($n$ $\sim$ 1--10$^3$ cm$^{-3}$). 
The blast wave in the western side of the SNR appears to be propagating into a lower density medium 
than in the east \citep{dick98}. Early studies estimated overall low metal abundances in N49, 
indicating that metal-rich ejecta might have been substantially intermixed with the interstellar 
medium (ISM) of the LMC \citep{danz85,russ90,hugh98}. 

The origin of N49 has been elusive (thermonuclear vs. core-collapse explosion). The interstellar
environment of N49, indicated by the presence of nearby molecular clouds \citep{bana97} and young 
stellar clusters \citep{klos04} and its location within an OB association \citep{chu88} generally 
support a core-collapse origin for N49 from a massive progenitor star. One model suggested that 
N49 is interacting with a dense shell of a Str\"omgren sphere created by a massive B-type progenitor 
\citep{shul85}. The positional coincidence of a soft $\gamma$-ray repeater (SGR) 0526--66 within 
the boundary of N49 \citep{clin82,roth94} may also suggest a core-collapse origin from a massive 
star, if this SGR is the compact remnant of the SN that created N49. However, the physical 
association between SNR N49 and SGR 0526--66 is uncertain \citep{gaen01,klos04,bade09}. Based on 
the {\it ASCA} data, Hughes et al.  (1998) found no evidence for surrounding ISM that had been 
modified by stellar winds from the massive progenitor. 

Nucleosynthesis studies of metal-rich ejecta in SNRs using spatially-resolved X-ray spectroscopy 
are  useful to constrain the progenitor's mass \citep[e.g.,][]{hugh03,park03b,park04,park07}. 
Although bright X-ray features in N49 are apparently dominated by emission from the shock-cloud 
interaction, previous {\it Chandra} observations have shown marginal evidence for enhanced emission 
lines and overabundant metal elements from a few sub-regions in the SNR \citep{park03a}. For instance, 
a promising ejecta candidate was a small protrusion beyond the SNR's southwestern boundary. This 
emission feature was suggested to be a candidate ejecta bullet, but metal overabundances were not 
conclusive because of poor photon statistics \citep{park03a}. Also, because N49 was detected 
significantly off-axis (by $\sim$6$\farcm$5) in the previous {\it Chandra} data (ObsID 1041) used 
by Park et al. (2003a), the point spread function (PSF) was substantially distorted, and the true 
morphology of this feature (point-like vs. extended) was uncertain. We note that there were three 
other on-axis {\it Chandra} observations of N49\footnote{ObsIDs 747, 1957, and 2515 with exposures 
of 40, 48, and 7 ks, respectively.}\citep{kulk03}. These previous observations were not useful to 
study the ejecta bullet candidate because of either the use of a CCD subarray or a very short 
exposure. Thus, a firm detection of metal-rich ejecta in N49 remains elusive.

Here we report on initial results from our new {\it Chandra} observation of N49. Our new {\it 
Chandra} data present $\sim$4 times deeper exposure (for the entire N49) than the data used by 
Park et al. (2003a), with an on-axis pointing to N49. With these new data, we clearly detect some 
metal-rich ejecta features in N49. Combining the new data with archival data, we also perform  
a detailed spectral analysis of SGR 0526--66. In this paper, we report on the results from our 
spectral analysis of SGR 0526--66 and the detected metal-rich ejecta features in N49. Our studies 
of complex shock-cloud interaction regions of N49 and the timing analysis of SGR 0526--66 will 
be presented in a subsequent paper. We describe the observations and the data reduction in 
Section~\ref{sec:obs}. The image and spectral analyses are presented in Section~\ref{sec:result}, 
and a discussion in Section~\ref{sec:disc}. A summary will be presented in Section~\ref{sec:sum}.

\section{\label{sec:obs} OBSERVATIONS \& DATA REDUCTION}

We observed N49 with the Advanced CCD Imaging Spectrometer (ACIS) on board {\it Chandra X-Ray 
Observatory} on 2009 July 18 -- September 19 (ObsIDs 10123, 10806, 10807, and 10808) during AO10.
We chose the ACIS-S3 chip to utilize the best sensitivity and energy resolution of the detector 
in the soft X-ray band. The pointing ($\alpha$$_{2000}$ = 05$^{h}$ 25$^{m}$ 58$^{s}$.8, 
$\delta$$_{2000}$ = $-$66$^{\circ}$ 05$'$ 00$\farcs$0) was roughly toward the geometrical center 
of N49. SGR 0526--66 and the ejecta bullet candidate are detected within $\sim$30$^{\prime\prime}$
and $\sim$40$^{\prime\prime}$ of the aim point, respectively. We used a 1/4 subarray of the 
ACIS-S3 to ensure low photon pileup ($\la$ 5\%) of SGR 0526--66, and to allow us to detect the 8 s 
pulsations from the SGR, while still obtaining full coverage of the entire SNR N49. We processed 
the raw event files following the standard data reduction methods using CIAO 4.2 and {\it Chandra} 
CALDB 4.3.0, which includes correction for the charge transfer inefficiency (CTI). We applied the 
standard data screening by status and grade ({\it ASCA} grades 02346). We performed this data 
reduction on individual observations. The overall light curve from the entire S3 chip for each 
ObsID showed no evidence of flaring particle background. Then, we merged four data sets into a 
single event file by re-projecting them onto ObsID 10807's tangential plane. After the data 
reduction, the total effective exposure is $\sim$108 ks. 

We also used two data sets (ObsIDs 747 and 1957) available in the {\it Chandra} archive\footnote{In 
the {\it Chandra} archive, there are two other data sets (ObsIDs 1041 and 2515) that detected N49 
and SGR 0526--66. However, these data are not useful for this work because of a large off-axis angle 
for N49 ($\sim$6$\farcm$5, ObsID 1041) and a short exposure ($\sim$7 ks, ObsID 2515).}. The pointing 
of these archival data was toward the position of SGR 0526--66 ($\alpha$$_{2000}$ = 05$^{h}$ 26$^{m}$ 
00$^{s}$.7, $\delta$$_{2000}$ = $-$66$^{\circ}$ 04$'$ 35$\farcs$0), and thus they provide a high 
resolution imaging of N49 as well as the SGR. Although these archival data cannot be used to study 
south-southwestern regions (including the ejecta bullet candidate) of N49 due to the use of 1/8 
subarray of the ACIS-S3, they covered east-northeastern parts of the SNR with a decent exposure 
(a total of $\sim$88 ks by combining ObsIDs 747 and 1957). Thus, these data are useful to study 
eastern regions of N49 and SGR 0526--66. We re-processed these two archival data sets in the same 
way as we did for our new observation data. Combining all these observations, our data present the 
deepest coverage with the high resolution X-ray imaging spectroscopy for SNR N49 ($\sim$108 ks in 
the south-west regions and $\sim$196 ks in the north-east regions) and SGR 0526--66 ($\sim$196 ks). 
Our new {\it Chandra} observations and archival data used in this work are summarized in 
Table~\ref{tbl:tab1}. 

\section{\label{sec:result} ANALYSIS \& RESULTS}

\subsection{\label{subsec:image} X-ray Images}

\begin{figure}[]
\figurenum{1}
\centerline{{\includegraphics[angle=0,width=\textwidth]{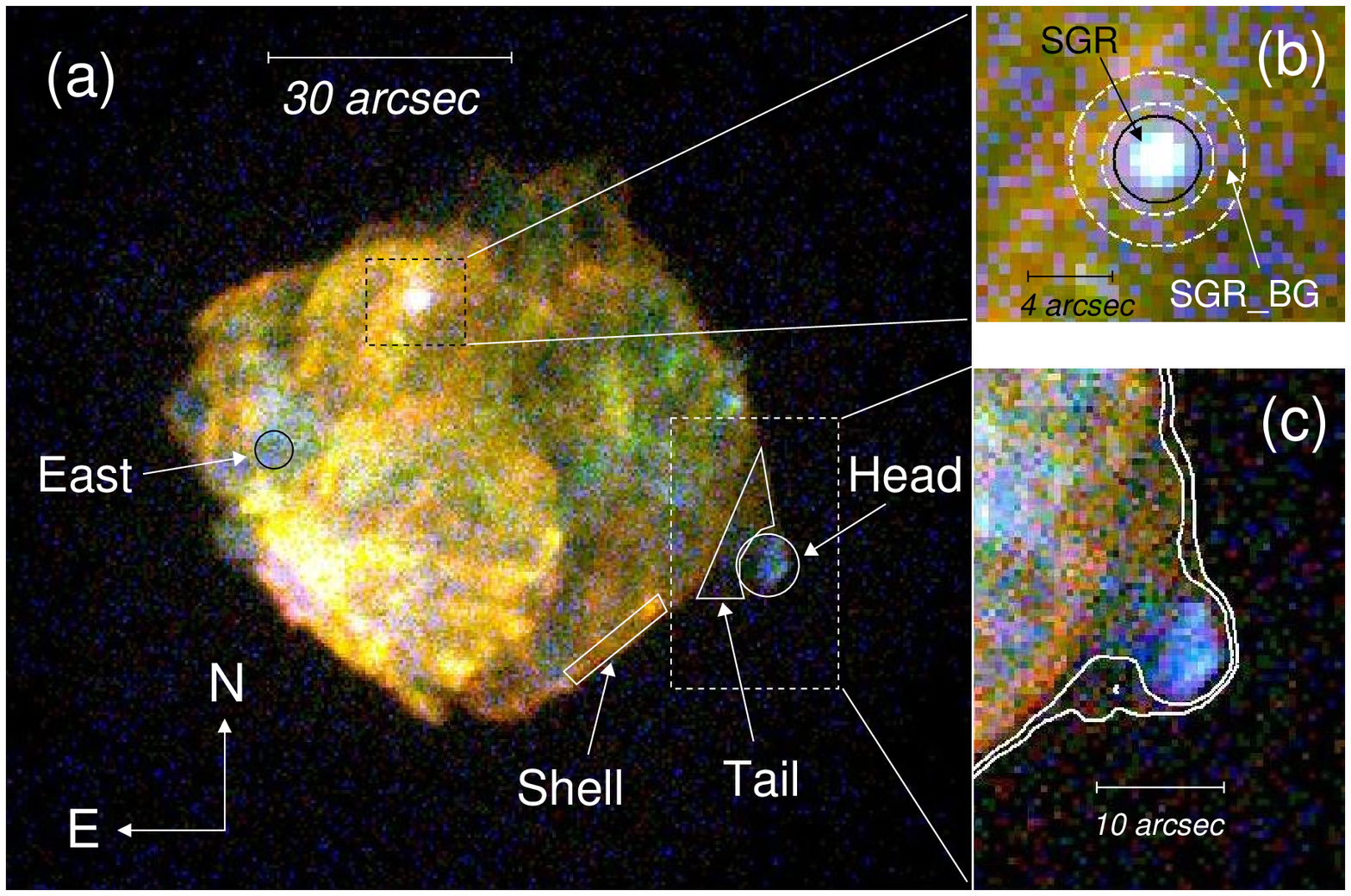}}}
\figcaption[]{(a) A 3-color ACIS images of N49 based on our 2009 data. Regions used for our 
spectral analysis are marked.  (b) A zoom-in image of the SGR 0526--66 region. The source and 
background regions used for the spectral analysis of SGR 0526--66 are shown. (c) A zoom-in of 
the southwestern region of N49 including the ejecta bullet. Contours for the outer boundary of 
N49 are overlaid. In (a), (b), and (c), Color codes are: red is 0.3--0.8 keV, green is 0.8--1.7 
keV, and blue is 1.7--7.0 keV band. The pixel size is 0$\farcs$5 in all panels.
\label{fig:fig1}}
\end{figure}

An X-ray color image of N49 shows complex asymmetric X-ray emission features (Figure~\ref{fig:fig1}a). 
These features have been partially revealed by previous {\it Chandra} data \citep{park03a,kulk03}. 
Our new deep {\it Chandra} image reveals all those fine X-ray structures throughout the entire SNR 
with substantially improved photon statistics and resolution (Figure~\ref{fig:fig1}a). The outer 
boundary of N49 appears as a thin shell radiating in soft X-rays (reddish in Figure~\ref{fig:fig1}a) 
that represents the blast wave sweeping through the general ambient medium of the LMC. A notable 
exception is the candidate ejecta bullet extending beyond the SNR shell in the southwest, which is 
distinctively blue in color (Figure~\ref{fig:fig1}c). Bright X-ray filaments in the eastern parts 
of the SNR are regions where the shock is interacting with dense clumpy clouds \citep[][and 
references therein]{park03a}. These shock-cloud interaction regions show a wide range in X-ray 
colors at angular scales down to several arcseconds. The western parts of N49 are relatively faint 
and X-ray emission is emphasized in green to blue while showing some red filamentary features as 
well (Figure~\ref{fig:fig1}a). This general color variation across the SNR appears to be related 
to the temperature variation caused by the overall interstellar density distribution, in which 
dense clouds are interacting with the shock mostly in the eastern half of the SNR \citep[e.g., 
][]{park03a}. 

\begin{figure}[]
\figurenum{2}
\centerline{\includegraphics[angle=0,width=\textwidth]{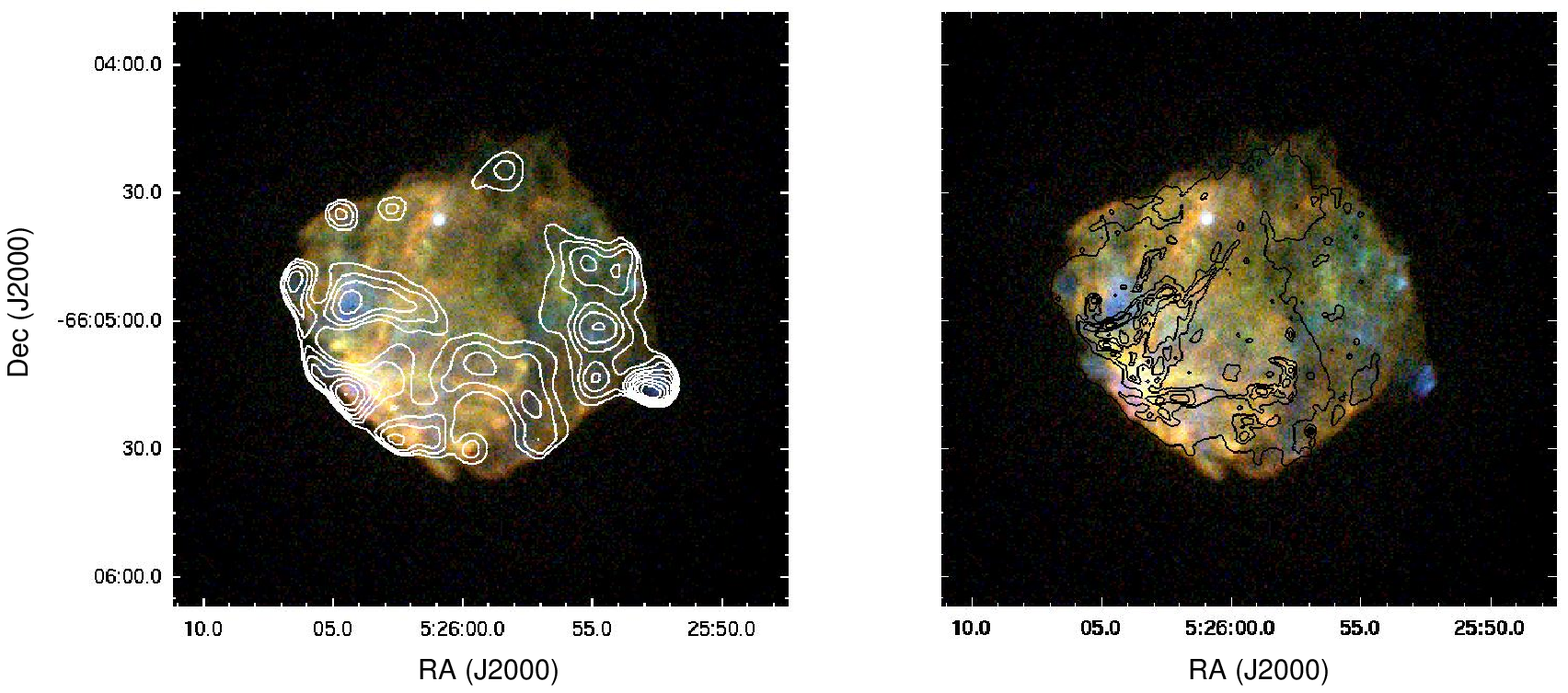}}
\figcaption[]{ (a) An X-ray 3-color image of N49 with contours of Si EW map overlaid.
Color codes are the same as those in Figure~\ref{fig:fig1}. The Si He$\alpha$ +
Ly$\alpha$ ($E$ = 1.75 -- 2.1 keV) EW has been calculated by the methods described in
Park et al. (2003a). (b) The same X-ray color image as in (a), overlaid with contours 
of an archival {\it Hubble Space Telescope} image.
\label{fig:fig2}}
\end{figure}

The spectrally-hard, blue protrusion extending beyond the main SNR shell in the southwestern 
boundary was proposed to be a candidate ejecta bullet based on the previous low-resolution {\it 
Chandra} data \citep{park03a}. If it is indeed an ejecta bullet, this feature is reminiscent of 
``shrapnel'' found in the Vela SNR \citep{asch95}, but nearly two orders of magnitude more 
distant. Our new {\it Chandra} data clearly resolve this feature. It is an extended feature 
with a head ($\sim$4$^{\prime\prime}$ in radius) and a tail that connects the head to the main 
SNR shell (the ``Head'' and ``Tail'' regions in Figure~\ref{fig:fig1}a, respectively). The blue 
color of the Head region is likely due to enhanced line emission in the hard X-ray band: e.g., 
the Si (He$\alpha$ + Ly$\alpha$) line {\it equivalent width} (EW) map\footnote{We created the Si 
line EW map following the method described in Park et al. (2003a). The 1.75--2.1 keV band was 
used to extract the Si line image. The upper and lower energy continuum images were created using
the 2.14--2.28 keV and 1.63--1.71 keV bands, respectively.} shows a strong enhancement in the 
Head region (Figure~\ref{fig:fig2}a), which suggests a metal-rich ejecta for the origin of this 
feature. In fact, our spectral analysis of the Head region shows highly overabundant Si and S 
(see Section~\ref{subsubsec:ejecta}). The Tail region appears to be composed of a central hard 
(blue) emission region surrounded by a soft (red), conical region (Figure~\ref{fig:fig1}c). It 
is likely the turbulent region behind the bullet, in which the metal-rich ejecta and shocked ISM 
are intermixed, possibly enclosed by the sides of a bow-shock produced by the bullet's supersonic 
motion. The bow shock, however, is not prominent ahead of the bullet, possibly due to unfavorable 
viewing geometry. The axis bisecting the opening angle ($\sim$87$^{\circ}$) is roughly pointing 
back to the geometric center of the SNR, but not exactly toward SGR 0526--66.

We noticed that there is a small region in the eastern part of the SNR (the ``East'' region in 
Figure~\ref{fig:fig1}a), which shows  similar characteristics to those of the Head region: 
i.e., this region stands out with a distinctively blue color in the bright eastern half of the 
SNR, and it is coincident with a strongly enhanced Si line EW (Figure~\ref{fig:fig2}a). The 
East region corresponds to an inter-cloud region (Figure~\ref{fig:fig2}b) in which a possibility 
of metal overabundance was suggested by our previous work \citep{park03a}. Our spectral analysis 
of this region indeed reveals its metal-rich ejecta nature (see Section~\ref{subsubsec:ejecta}).

\subsection{\label{subsec:spec} X-ray Spectra} 

Based on results from our image analysis, we extracted spectra from several regions that 
appear to characteristically represent distinctive features: i.e., the swept-up interstellar 
medium (the ``Shell'' and ``SGR\_BG'' regions), shocked metal-rich ejecta (the ``Head'' and 
``East'' regions), and the connecting region between the ejecta bullet and the main SNR shell 
(the ``Tail'' region). These regions are marked in Figures~\ref{fig:fig1}a \& \ref{fig:fig1}b. 
We also extracted the spectrum from the quiescent X-ray counterpart of SGR 0526--66 (the ``SGR'' 
region in Figure~\ref{fig:fig1}b). We present the spectral analysis of these swept-up medium, 
ejecta features, and 0526--66 in Sections~\ref{subsubsec:ism}, \ref{subsubsec:ejecta}, and 
\ref{subsubsec:sgr}, respectively.

\subsubsection{\label{subsubsec:ism} N49: Swept-Up Interstellar Medium}

The brightest X-ray emission in the eastern parts of N49 originates from the shock interaction 
with clumpy clouds with a density gradient spanning a few orders of magnitudes \citep{vanc92}. 
X-ray emission from such a clumpy medium would represent a complex mixture of gas temperatures 
and ionization states.  Such a plasma condition is likely localized in strong shock-cloud 
interaction regions.  While the western regions do not show evidence of shock interaction with 
dense clumpy clouds, they also show some significant X-ray color variation (Figure~\ref{fig:fig1}a). 
For those regions, it is not straightforward to characterize the X-ray emission spectrum of the 
general ambient medium shocked by the blast wave. Thus, to study the nature of the general swept-up 
ISM and the SNR's overall dynamics, we extracted the X-ray spectrum from a region in the outermost 
rim of the SNR in the southern boundary (the ``Shell'' region in Figure~\ref{fig:fig1}a) in which 
no complex spectral and/or spatial substructures are seen. We also used a small annular region 
surrounding SGR 0526--66 (the ``SGR\_BG'' region in Figure~\ref{fig:fig1}b) to characterize 
background emission for SGR 0526--66. Since the SGR\_BG region is in the {\it transition} area 
between the bright eastern and faint western parts of the SNR, it may also effectively show an 
average plasma condition of the complex shock structures across the SNR. 

For our spectral analysis, we re-binned the observed spectra to contain a minimum of 20 
counts per energy channel. We fit the X-ray spectrum with a non-equilibrium ionization (NEI) 
state plane-shock model \citep[vpshock model in conjunction with the NEI version 2.0 in the 
XSPEC software,][]{bork01} that is based on ATOMDB \citep{smit01}. We used  an augmented 
version\footnote{This augmented NEI model has been provided by K. Borkowski. A relevant 
discussion on this modeling issue can be found in Badenes et al. (2006).} of this atomic 
database to include inner-shell processes (e.g., lines from Li-like ions) and updates of 
the Fe L-shell lines, whose effects are substantial in under-ionized plasma, but are not 
incorporated in the standard XSPEC NEI version 2.0. We fixed the Galactic column at 
$N_{\rm H,Gal}$ = 6 $\times$ 10$^{20}$ cm$^{-2}$ toward N49 \citep{dick90}. We fit the 
foreground column by the LMC ($N_{\rm H,LMC}$) assuming the LMC abundances available in 
the literature \citep[He = 0.89, C = 0.30, N = 0.12, O = 0.26, Ne = 0.33, Na = 0.30, Mg 
= 0.32, Al = 0.30, Si = 0.30, S = 0.31, Cl = 0.31, Ar = 0.54, Ca = 0.34, Cr = 0.61, Fe = 
0.36, Co = 0.30, and Ni = 0.62, ][]{russ92,hugh98}. Hereafter, elemental abundances are 
with respect to Solar \citep{ande89}.

\begin{figure}[]
\figurenum{3}
\centerline{\includegraphics[angle=0,width=\textwidth]{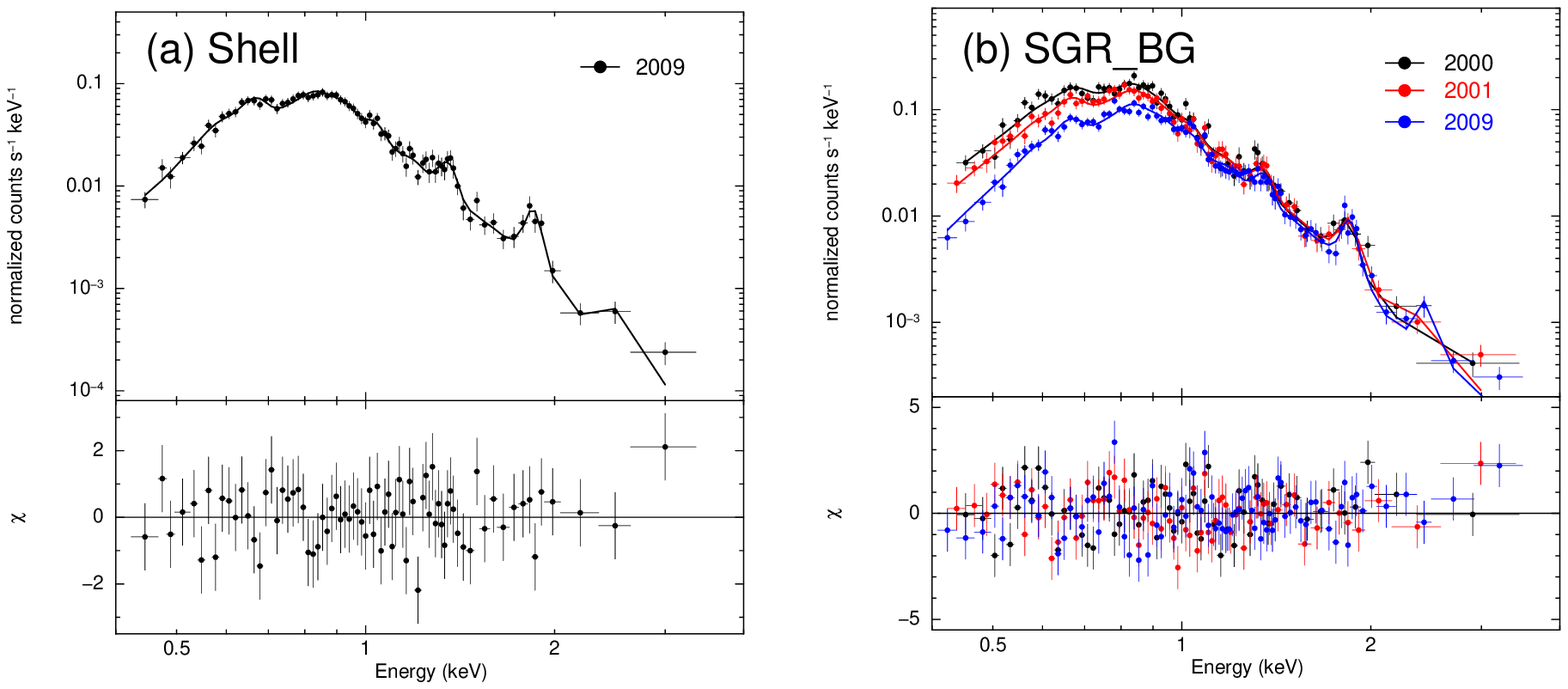}}
\figcaption[]{(a) X-ray spectrum of ``Shell'' region. (b) X-ray spectrum
of ``SGR\_BG'' region. In (a) and (b), the best-fit plane-shock model
is overlaid. The lower panel is the residuals from the best-fit model. 
\label{fig:fig3}}
\end{figure}

The southern boundary of N49 was not covered by ObsIDs 747 and 1957 because of the use 
of a 1/8 subarray. Thus, we used our new {\it Chandra} data to extract the spectrum from 
the Shell region. The background was subtracted using the spectrum extracted from two 
circular source-free regions (with a radius of 8$^{\prime\prime}$) beyond the south-western
boundary of the SNR. The Shell region spectrum contains significant photon statistics
of $\sim$4600 counts, and can be well fitted by the NEI plane-shock model ($\chi^2$/n = 
49.1/68 with the electron temperature $kT$ = 0.57 keV and the ionization timescale $n_et$ 
= 6.35 $\times$ 10$^{11}$ cm$^{-3}$ s, Figure~\ref{fig:fig3}a). The foreground column by 
the LMC is estimated to be $N_{\rm H,LMC}$ $\sim$ 0.9 $\times$ 10$^{21}$ cm$^{-2}$. The 
best-fit metal abundances are generally consistent with the LMC values. Results from this 
spectral model fit are summarized in Tables~\ref{tbl:tab2} \& \ref{tbl:tab3}.

The SGR\_BG region was observed by ObsIDs 747 and 1957 as well as our new {\it Chandra}
observations, and thus, we used all of those data to extract the spectrum from this region.
The total photon statistics for this region is $\sim$13600 counts. The background was 
subtracted using the spectrum extracted from a circular source-free region (with a radius of 
8$^{\prime\prime}$) beyond the northern boundary of the SNR. We initially fit these three 
data sets simultaneously with all fitted parameters tied with each other. Although the overall 
fit may be acceptable ($\chi^2$/n = 1.28), there appears to be a small systematic offset in the 
normalization at $E$ $\la$ 1 keV between data taken in 2000/2001 (ObsIDs 747 and 1957) and in 
2009 (ObsIDs 10123+10806+10807+10808). Considering the fact that the time separation between 
2000/2001 and 2009 observations is substantial and that the effect appears to be emphasized in 
the soft band ($E$ $\la$ 1 keV), this small discrepancy is likely related to the calibration 
effect by the time-dependent quantum efficiency degradation in the ACIS. Since the quantum 
efficiency degradation affects the spectral modeling as if there is an ``extra'' foreground 
absorption, we attempted the model fit with $N_{\rm H, LMC}$ untied among three spectra. The 
best-fit $N_{\rm H, LMC}$ values are the same between 2000 and 2001 data, while it is $\sim$25\% 
larger for the 2009 data. Thus, we repeated the spectral model fit with $N_{\rm H, LMC}$ tied 
only between 2000 and 2001 data, and with that for 2009 data allowed to vary freely. The fit 
improved somewhat ($\chi^2$/n = 1.19). The best-fit columns are $N_{\rm H, LMC}$ = 
1.39$^{+0.46}_{-0.47}$ $\times$ 10$^{21}$ cm$^{-2}$ and 1.76$^{+0.43}_{-0.38}$ $\times$ 
10$^{21}$ cm$^{-2}$ for the 2000/2001 and 2009 data, respectively. While showing a marginal 
calibration effect, these values are consistent within statistical uncertainties (hereafter, 
errors are with a 90\% confidence level). In the following discussion, we assume the average 
value of $N_{\rm H, LMC}$ = 1.58$^{+0.45}_{-0.44}$ $\times$ 10$^{21}$ cm$^{-2}$. The best-fit 
spectral parameters for the SGR\_BG region are $kT$ $\sim$ 0.56 keV and $n_et$ $\sim$ 9.65 
$\times$ 10$^{11}$ cm$^{-3}$ s, all of which are consistent with those for the Shell region 
within uncertainties. The best-fit metal abundances are also generally consistent with those 
from the Shell region. These model parameters are summarized in Tables~\ref{tbl:tab2} \& 
\ref{tbl:tab3}.

\subsubsection{\label{subsubsec:ejecta} N49: Metal-Rich Ejecta}

\begin{figure}[]
\figurenum{4}
\centerline{\includegraphics[angle=0,width=\textwidth]{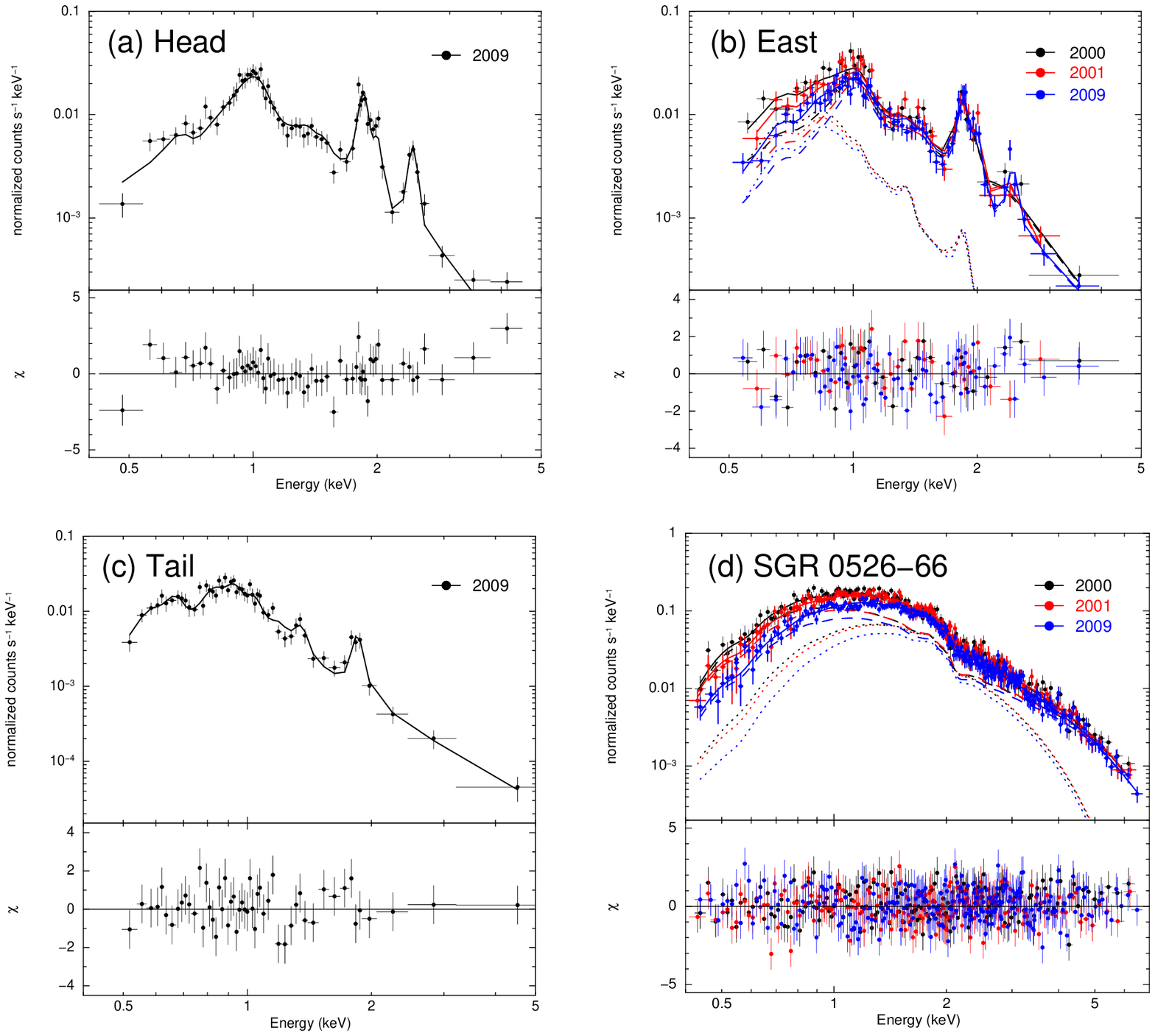}}
\figcaption[]{X-ray spectrum of (a) ``Head'', (b) ``East'', (c) ``Tail'', and (d) SGR 0526--66. 
In (a), (b), and (c), the best-fit plane-shock model is overlaid (solid curve). In (d), the 
best-fit BB+PL model is overlaid (solid curve). In (b), dashed and dotted curves show model 
components for the ejecta and the superposed emission from the shocked cloud, respectively. 
In (d), dashed and dotted curves show the best-fit PL and BB model components, respectively. 
The lower panel is the residuals from the best-fit model. 
\label{fig:fig4}}
\end{figure}

Our new {\it Chandra} data clearly resolve the extended nature of the ejecta fragment bullet 
(the Head region) and the trailing tail-like hot gas (the Tail region, Figure~\ref{fig:fig1}c). 
Because the bullet region was not covered by the 2000/2001 data, we used only 2009 data for the 
spectral analysis of this feature. We extracted $\sim$1800 counts from the Head region. In 
contrast to those for the swept-up medium, the X-ray spectrum of the Head region shows remarkably 
enhanced emission lines from highly ionized Si and S (Figure~\ref{fig:fig4}a), which is consistent 
with the corresponding bright Si line EW image (Figure~\ref{fig:fig2}a). We fitted this spectrum 
with an NEI plane-shock model following the method described in Section~\ref{subsubsec:ism}. 
The background was subtracted using the same spectrum as we used for the Shell region. We 
initially fixed metal abundances at the best-fit values for the Shell region (Table~\ref{tbl:tab3}), 
which resulted in a statistically unacceptable fit ($\chi^2$/n = 3.3) primarily because of the 
strongly enhanced Si and S lines. Varying Si and S abundances provides an acceptable fit ($\chi^2$/n 
= 67.0/59, Figure~\ref{fig:fig4}a). The best-fit LMC column ($N_{\rm H,LMC}$ $\sim$ 1.3 $\times$ 
10$^{21}$ cm$^{-2}$) is consistent with those estimated from the Shell and SGR\_BG regions. The 
electron temperature ($kT$ $\sim$ 1 keV) and the ionization timescale ($n_et$ $>$ 2 $\times$ 
10$^{12}$ cm$^{-3}$ s) are significantly higher than those from swept-up medium. The Si ($\sim$2.3) 
and S ($\sim$3.2) abundances are an order of magnitude larger than the LMC values, firmly 
establishing the metal-rich ejecta nature of the bullet. 

Varying other elemental abundances (O, Ne, Mg, and Fe) somewhat improves the fit ($\chi^2$/n = 
49.3/55, but the corresponding F-probability is marginal [$\sim$0.002]), and the best-fit electron 
temperature is very high ($kT$ $\sim$ 4 keV). In addition to the overabundant Si and S, the best-fit 
abundance appears to be moderately enhanced for O ($\sim$0.7), Ne ($\sim$2.1), and Mg ($\sim$1.0). 
However, these abundances are not well constrained (with a factor of $\ga$2 uncertainties), and thus 
the overabundance for these elements is not convincing based on the current data. We conclude that 
the main differences in the fitted parameters resulting from varying metal abundances other than 
Si and S (i.e., the high $kT$ $\sim$ 4 keV and possibly enhanced abundances for O, Ne, and Mg) 
are not compelling.  Otherwise, it does not affect our discussion of the ejecta features (see 
Section~\ref{subsec:disc_ejecta}). Thus, we hereafter discuss the nature of the bullet based on 
our spectral model fit with only Si and S abundances varied. The results from the Head region model 
fit are summarized in Tables~\ref{tbl:tab2} \& \ref{tbl:tab3}.

The Tail region is faint, and we extracted $\sim$1500 counts in this region. We fit this spectrum 
with an NEI plane shock model $\chi^2$/n = 44.9/44, Figure~\ref{fig:fig4}c). The Tail region is 
fitted by the X-ray spectrum from a hot gas ($kT$ $\sim$ 2 keV) in a significantly under-ionized 
state ($n_et$ $\sim$ 7 $\times$ 10$^{10}$ cm$^{-3}$ s). The best-fit elemental abundances are 
consistent with those estimated from the Shell and SGR\_BG regions. We note that the Tail region 
shows substructures of the spectrally-hard (blue) {\it interior} surrounded by soft (red) outer 
layer (Figure~\ref{fig:fig1}c). It suggests that the Tail region may be a mixture of metal-rich 
ejecta and shocked swept-up medium. Thus, we attempted an alternative spectral model, assuming two 
shock components, a soft component for the swept-up ISM and a hard component for the metal-rich 
ejecta.  The spectral parameters, except for the normalization, and metal abundances for the ejecta 
component were fixed at the best-fit values estimated from the Head region. The metal abundances 
for the soft component were fixed at the values estimated from the Shell region. The electron 
temperature, ionization timescale, and the normalization parameters for the soft component were 
varied freely.  The observed Tail spectrum can be equally fitted by this two-component model 
($\chi^2$/n = 52.4/49). In this model fit, the metal-rich ejecta component contributes $\sim$15\% 
of the total observed X-ray flux in the 0.5--10 keV band. These one- and two-component models are 
not statistically distinguished with the current data. Results from these Tail region spectral 
model fits are summarized in Tables~\ref{tbl:tab2} \& \ref{tbl:tab3}.

The East region shows a distinctively blue color (Figure~\ref{fig:fig1}a) with a strong enhancement 
in the Si line EW image (Figure~\ref{fig:fig2}a), which resembles the spectral properties of the 
Head region of the bullet.  This region corresponds to a hot inter-cloud region \citep{park03a}. 
This region may also be a candidate ejecta feature, probably isolated from the complex emission 
from shocked dense clumpy clouds (Figure~\ref{fig:fig2}b). Since this region was covered by both 
the 2000/2001 and 2009 data, we used all of these data to extract the X-ray spectrum. We 
extracted the X-ray spectrum from a small circular region with a radius of 2$^{\prime\prime}$, 
in which the {\it Chandra} data allowed us total photon statistics of $\sim$3600 counts. The 
background was subtracted using the spectrum extracted from two circular source-free regions (with 
a radius of 8$^{\prime\prime}$) beyond the eastern boundary of the SNR.  We fit the observed spectrum 
with the NEI plane shock model.  Because enhanced Si-K and S-K lines are evident, we varied Si and S 
abundances while fixing other abundances at average values estimated from the Shell and SGR\_BG 
regions.  The fitted parameters are $N_{\rm H,LMC}$ $\sim$ 0.8 $\times$ 10$^{21}$ cm$^{-2}$, $kT$ 
$\sim$ 1.05 keV, and $n_et$ $>$ 52.4 $\times$ 10$^{11}$ cm$^{-3}$ s. The estimated Si and S abundances 
are high (Si $\sim$ 1.8 and S $\sim$ 1.4). While the fit may be acceptable ($\chi^2$/n = 173.4/134), 
there appear to be some systematic residuals at $E$ $\la$ 1 keV. Based on the results from our spectral 
analysis of the ejecta (the Head region) and the swept-up ISM (the Shell and SGR\_BG regions) regions, 
we consider that the East region emission is a superposition of the spectrally-hard metal-rich ejecta 
and the projected soft swept-up ISM components. To account for the contribution of soft X-ray emission 
from the shocked ISM in the observed spectrum of the East region, we added an NEI model component for 
which we fixed the electron temperature, ionization timescale, and metal abundances at the average 
values estimated from the Shell and SGR\_BG regions. Only the normalization parameter was varied for 
this component.  This model fit significantly improves ($\chi^2$/n = 144.4/133) without affecting the 
best-fit parameters for the hard, ejecta component. The only noticeable change is a larger value for 
the best-fit $N_{\rm H,LMC}$ $\sim$ 3 $\times$ 10$^{21}$ cm$^{-2}$. This larger estimate for $N_{\rm 
H,LMC}$ appears to be consistent with the presence of dense molecular clouds toward eastern parts 
of the SNR. The assumed soft component from the swept-up ISM contributes $\sim$20\% of the total 
observed X-ray flux in the 0.5--5 keV band. The results from this fit are summarized in 
Table~\ref{tbl:tab2} \& \ref{tbl:tab3}.

\subsubsection{\label{subsubsec:sgr} SGR 0526--66}

The X-ray spectrum of SGR 0526--66 is extracted from a circular region of 2$^{\prime\prime}$
radius (Figure~\ref{fig:fig1}b). The background spectrum is characterized by emission from an
annular region (2$^{\prime\prime}$ in thickness) around the source (the SGR\_BG region, 
Figure~\ref{fig:fig1}b) to average the complex diffuse emission from SNR N49 projected against 
the SGR. Since archival {\it Chandra} data (ObsIDs 747 and 1957) were pointed at the SGR with 
a decent exposure (Table~\ref{tbl:tab1}), we used these archival data as well as our new data 
for the spectral analysis of SGR 0526--66. We fit these three spectra simultaneously. Just like 
our spectral analysis of N49, we assumed that the SGR spectrum is absorbed by both Galactic 
(fixed at $N_{\rm H,Gal}$ = 6 $\times$ 10$^{20}$ cm$^{-2}$) and LMC columns ($N_{\rm H,LMC}$). 

Initially, we fit the SGR spectrum with a single power law (PL) in which we tied all fitted 
parameters between the three sets of the spectrum. The best-fit model from this fit is not
acceptable ($\chi$$^2$/$n$ $\sim$ 1.8). The best-fit model shows significant residuals at
$E$ $\ga$ 0.8 keV, because it underestimates the overall flux level for the 2000/2001 data 
while it overestimates the flux level for the 2009 data. Thus, we untied the normalization 
parameter between the 2000/2001 and 2009 data, and then repeated the model fit. The fit improved 
($\chi$$^2$/$n$ $\sim$ 1.5) with a $\sim$15\% lower normalization for the 2009 data than that 
for the 2000/2001 data. Although statistically improved by removing the normalization offset
between the 2000/2001 and 2009 data, we do not accept this fit because systematic residuals 
are evident over the entire bandpass with a relatively high $\chi$$^2$/$n$ $\sim$ 1.5. 
Alternatively, we attempted a single blackbody (BB) model fit. The best-fit models are 
statistically unacceptable ($\chi$$^2$/$n$ $\sim$ 2.2 and 2.5 with the untied and tied 
normalization parameter between the 2000/2001 and 2009 data, respectively).

Kulkarni et al. (2003) showed that a PL+BB model was needed to adequately describe the quiescent
X-ray emission from SGR 0526--66. On the other hand, Park et al. (2003a) argued that there could 
have been a considerable contamination from the surrounding soft thermal X-ray emission in the 
SGR spectrum presented by Kulkarni et al. (2003), and that the need for a BB component was not 
conclusive. Our SGR spectrum with $\sim$4 times higher photon statistics than that used by Park 
et al. (2003a) indicates that a PL model cannot describe the observed SGR spectrum, supporting 
the conclusion by Kulkarni et al. (2003). Thus, we fit the SGR spectrum with a PL+BB model. 
We repeated the general steps as we did with a single PL model fits: i.e., we first tied all 
fitted parameters between the 2000/2001 and 2009 data sets, and then untied them one at a time 
(i.e., BB temperature [$kT_{\rm BB}$], PL photon index [$\Gamma$], and normalizations for the BB 
and PL components). The best-fit values for $kT_{\rm BB}$ and $\Gamma$ are consistent within 
uncertainties between the 2000/2001 and 2009 data. The normalization for the BB and/or PL 
components are fitted to be $\sim$10--30\% lower for the 2009 data than they are for the 2000/2001 
data. The fit improvement due to this normalization change is statistically significant, and the 
fit is acceptable ($\chi$$^2$/$n$ = 577.4/582). The normalization change appears to be more 
effective for the PL component than for the BB, probably because the PL component covers a broader 
range of the observed spectrum than the BB component (The PL component contributes $\sim$70\% of 
the total observed flux in the 0.5--10 keV band). 

While the X-ray spectra of SGRs and anomalous X-ray pulsars (AXPs) are usually fitted with a 
BB+PL model \citep[e.g., ][]{mere07}, double BB models (BB+BB) have also been suggested to fit 
some AXPs \citep[e.g., ][]{halp05}. A recent work showed that a BB+BB model fit can successfully 
reproduce the observed {\it Chandra} spectrum (taken in 2000 and 2001) of SGR 0526--66 \citep{naka09}. 
Thus, we attempted to fit our SGR spectrum with a BB+BB model. We tied BB temperatures between 
2000/2001 and 2009 data since we found no evidence of change in the temperatures. Just like our 
BB+PL modeling, we varied normalizations freely between 2000/2001 and 2009. The fit is statistically 
good ($\chi$$^2$/$n$ = 616.4/582). Although statistical uncertainties are relatively large, a 
$\sim$15\% decrease in the best-fit value of the normalization parameters are suggested for both BB 
components, which is similar to the results from the BB+PL model fit.  

Both of the BB+PL and BB+BB model fits suggest that the overall X-ray flux of SGR 0526--66 in the 
0.5--10 keV band has been reduced by $\sim$15\% in 2009 compared with that in 2000/2001. We note 
that we used a 1/4 subarray of the ACIS-S3 in our 2009 observations, while a 1/8 subarray was used 
in 2000 and 2001. Based on the event grade distribution, we found that the photon pileup effect is 
similarly small in both data sets ($\sim$2\% in 2000 and 2001 vs. $\sim$4\% in 2009). Nonetheless,
assuming the X-ray flux observed in 2000/2001, we estimated the pileup effect on the X-ray flux 
measurements in 2009 data using the Portable Interactive Multi-Mission Simulator (PIMMS). Our 
PIMMS simulations predict that the pileup effect in 2009 data would reduce the 0.5--10 keV band 
X-ray flux in 2009 by $\la$5\% compared with that measured in 2000/2001. We also tested the pileup
effect by applying an ACIS pileup model \citep{davi01} for our spectral model fits. This model 
indicated that the 0.5--10 keV flux estimates with the 2009 data are probably affected by $\la$5\% 
due to the use of 1/4-subarray. Thus, although the flux change between 2000/2001 and 2009 is
partially caused by pileup effect, the pileup is not responsible for the entire flux change.  
The apparent flux change is unlikely due to uncertainties in the detector calibration for the
time-dependent quantum efficiency degradation, because there is no evidence for discrepancy in
the fitted normalization parameters that is emphasized in the soft band ($E$ $\la$ 1 keV) between
2000/2001 and 2009 data. Thus, while follow-up {\it Chandra} observations are required to draw a 
firm conclusion on the nature of the long-term X-ray light curve of 0526--66, we tentatively 
conclude that the discrepancy in the observed X-ray flux of SGR 0526--66 between 2000/2001 and 
2009 is probably real rather than an artifact due to the photon pile-up and/or an inaccurate 
calibration of the time-dependent quantum efficiency degradation between two epochs. The SGR 
spectrum and the best-fit BB+PL model are presented in Figure~\ref{fig:fig4}d. Results from 
the best-fit BB+PL and BB+BB model fits are summarized in Table~\ref{tbl:tab4} and \ref{tbl:tab5}, 
respectively. 

\section{\label{sec:disc} DISCUSSION}

\subsection{\label{subsec:disc_ejecta} N49}

Based on the volume emission measure ($EM$) estimated from the spectral fit of the Shell region, 
we calculate the post-shock electron density ($n_e$) behind the blast wave. We assumed an emission 
volume of $V$ $\sim$ 9.4 $\times$ 10$^{55}$ cm$^3$ for a 3$^{\prime\prime}$ $\times$ 
18$^{\prime\prime}$ region (corresponding to $\sim$0.7 pc $\times$ 4.4 pc, hereafter we assume 
$d$ = 50 kpc for the LMC) with a $\sim$1 pc path-length along the line of sight. We also assumed 
$n_e$ = 1.2 $n_H$ for a mean charge state with normal composition. The best-fit $EM$ (= 6.75 
$\times$ 10$^{57}$ cm$^{-3}$) implies $n_e$ $\sim$ 9.3$f^{-{1\over2}}$ cm$^{-3}$ and $n_H$ 
$\sim$ 7.7$f^{-{1\over2}}$ cm$^{-3}$, where $f$ is the volume filling factor of the X-ray emitting
gas. The pre-shock H density is then $n_0$ $\sim$ 1.9$f^{-{1\over2}}$ cm$^{-3}$ for a strong 
adiabatic shock where $n_H$ = 4$n_0$. Assuming an adiabatic shock in electron-ion temperature 
equipartition, the gas temperature is related with the shock velocity ($v$) as $T$ = 
3$\bar{m}$$v^2$/16$k$ where $k$ is the Boltzmann constant and $\bar{m}$ $\sim$ 0.6 $m_p$ is 
the mean molecular weight with the proton mass $m_p$ = 1.67 $\times$ 10$^{-24}$ g. For the gas 
temperature of $kT$ = 0.57 keV, the shock velocity $v$ $\sim$ 700 km s$^{-1}$ is estimated. 
Thus, for the radius of $\sim$8.5 pc (assuming the angular distance of $\sim$35$^{\prime\prime}$ 
between the Shell region and the geometric center of N49), we estimate the Sedov age of the SNR, 
$\tau_{\rm Sed}$ $\sim$ 4800 yr. The estimated $\tau_{\rm Sed}$ and $n_0$ imply an explosion 
energy of $E_0$ $\sim$ 1.8$f^{-{1\over2}}$ $\times$ 10$^{51}$ erg. The estimated Sedov age is 
$\sim$30\% lower than our previous estimate \citep[$\sim$6600 yr,][]{park03a} for which we 
assumed $n_0$ = 0.9 cm$^{-3}$ based on optical data \citep{vanc92} and the canonical value of 
the SN explosion energy ($E_0$ = 1 $\times$ 10$^{51}$ erg). Also, the previous estimate was based 
on a more complex spectral modeling of multi-temperature plasma in the eastern regions of the SNR 
for which an accurate estimate of the distance from the SNR center was not feasible. Thus, we 
conclude that our new estimate of the SNR age is more reliable, while being in plausible agreement 
with the previous estimate by Park et al. (2003a).

The presence of the ejecta bullet beyond the southwestern shell of N49 is conclusively established 
by our new {\it Chandra} observation. The on-axis ACIS image clearly resolves the bullet into a 
head and a tail extending back to the main shell of the SNR. The head region itself is an extended 
feature ($\sim$4$^{\prime\prime}$ in radius) with an enhanced intensity toward the inferred 
direction of motion (Figure~\ref{fig:fig1}c). The X-ray color of the head is distinctively blue in 
contrast to reddish color of the overall SNR shell. The foreground column ($N_{\rm H,LMC}$ $\sim$ 
1.3 $\times$ 10$^{21}$ cm$^{-2}$) toward the head is consistent with those to N49's main shell and 
the bullet's tail region, which supports that the blue color is intrinsic for the head rather than 
being caused by a significantly larger foreground absorption (in which case the head would probably 
be a background source). In fact, the observed X-ray spectrum (Figure~\ref{fig:fig4}a) shows that 
the hardness of the head is due to highly enhanced line emission from He- and H-like Si and 
He-like S ions. The estimated abundances for Si ($\sim$2.3) and S ($\sim$3.2) are an order of 
magnitude larger than the LMC values. We note that Park et al. (2003a) suggested a possibility of 
the enhanced abundance for O in addition to Si and S in the bullet. However, the overabundance for 
O is not confirmed by our new data. Nonetheless, the highly overabundant Si and S firmly establish 
the identification of the bullet as metal-rich stellar fragment ejected from the deep interior of 
the progenitor star. 

The East region shows nearly identical spectral characteristics to the Head region: i.e., a 
distinctively blue color compared with surrounding regions and strongly enhanced Si and S lines. 
Just like the Head region, significantly enhanced Si and S abundances are estimated, but 
overabundances for other metal species are not required to describe the observed spectrum. The 
foreground absorption for the East region appears to be larger than other regions of N49, and 
it is probably caused by nearby dense molecular clouds which are interacting with the SNR in 
the eastern regions \citep[e.g., ][]{bana97,otsu10}. If the {\it excess} column ($\Delta$$N_{\rm 
H,LMC}$ $\sim$ 2 $\times$ 10$^{21}$ cm$^{-2}$ compared with the average $N_{\rm H,LMC}$ $\sim$ 1.2 
$\times$ 10$^{21}$ cm$^{-2}$ estimated for other regions of N49, Table~\ref{tbl:tab2}) originates
from these nearby molecular clouds, the corresponding average cloud density of $n_H$ $\sim$ 90 
cm$^{-3}$ is implied for the overall cloud size of $R$ $\sim$7 pc (the size estimated by Banas 
et al. 1997). This average density is in plausible agreement with the pre-shock density range 
of the clumpy clouds ($n_0$ $\sim$ 20--940 cm$^{-3}$, Vancura et al. 1992) with which N49 is 
interacting, while it is significantly larger than the value we estimated for the Shell region 
($n_0$ $\sim$ 1.9 cm$^{-3}$) where the shock is propagating into the low density ambient medium. 

Based on the Si and S abundances and volume emission measures of these metal-rich ejecta features, 
we estimate the Si to S ejecta mass ratio. The observed spectrum of the Head and East regions shows 
that He- and H-like ionization states dominate for Si, while S ions are primarily in He-like state. 
Thus, for simplicity, we assumed a ``pure'' ejecta case with electron to ion density ratios of 
$n_{e,{\rm Si}}$ $\approx$ 12.5 $n_{\rm Si}$ and $n_{e,{\rm S}}$ $\approx$ 14 $n_{\rm S}$. For 
dominant isotopes of $^{28}$Si and $^{32}$S, the measured Si and S abundances imply the ejecta 
mass ratio $M_{\rm Si}$/$M_{\rm S}$ $\sim$ 1.2 and 1.6 ($V_{\rm Si}$/$V_{\rm S}$)$^{1\over2}$, 
where $V_{\rm Si}$ and $V_{\rm S}$ are the volume of Si- and S-rich hot ejecta gas, respectively. 
Assuming $V_{\rm Si}$ $\approx$ $V_{\rm S}$, the estimated mass ratios are $M_{\rm Si}$/$M_{\rm S}$ 
$\sim$1.2 for the Head region and $\sim$1.6 for the East region. The mixture of H in metal-rich 
ejecta may not significantly affect our mass ratio estimates as long as the ISM mixture rate is 
similar between the Si and S ejecta. We compare these $M_{\rm Si}$/$M_{\rm S}$ with standard SN 
nucleosynthesis models. Our estimated $M_{\rm Si}$/$M_{\rm S}$ appears to be smaller than those 
for core-collapse SN models in which $M_{\rm Si}$/$M_{\rm S}$ typically ranges $\sim$2--4 depending 
on the progenitor's mass \citep[e.g.,][]{nomo97a,raus02,limo03}. The Si to S mass ratio for Type Ia 
SN models ($M_{\rm Si}$/$M_{\rm S}$ $\sim$ 1.5 -- 1.8, e.g., Nomoto et al. [1997b]; Iwamoto et al. 
[1999]) are generally smaller than core-collapse cases, which are closer to our estimates for N49. 
The lack of evidence for O-rich ejecta in the Head and East regions is generally suggestive of a 
Type Ia origin as well. If we take $\sim$0.3 (Table~\ref{tbl:tab3}) as an {\it upper limit} for 
the O abundance of the SN nucleosynthesis in N49, $M_{\rm O}$/$M_{\rm Si}$ $<$ 1.4 can be inferred 
for the Head region. This limit appears to be consistent with Type Ia models ($M_{\rm O}$/$M_{\rm 
Si}$ $\la$ 1, e.g., Nomoto et al. [1997b]; Iwamoto et al. [1999]), while being smaller than  
those for core-collapse models ($M_{\rm O}$/$M_{\rm Si}$ $>$ 2, e.g., Nomoto et al. [1997a]; 
Limongi \& Chieffi [2003]). 

On the other hand, the Head and East regions do not show evidence for overabundant Fe, which 
is usually considered as an iconic feature for Type Ia SNRs. The lack of Fe-rich ejecta is 
problematic for a Type Ia interpretation, and it would rather support a core-collapse origin. 
Also, the suggested Type Ia origin for N49 is inconsistent with its interstellar environment 
with recent star-forming regions and nearby molecular clouds, which rather suggests a massive 
progenitor for N49 \citep[e.g.,][]{chu88,bana97,klos04,bade09}. If N49 is a remnant of a 
core-collapse explosion from a massive progenitor rather than a Type Ia, the Si-rich nature 
of the Head and East regions may be generally considered to be explosive O-burning or incomplete 
Si-burning products from deep inside of the core-collapse SN. In fact, Si-rich ejecta knots have 
been detected in core-collapse SNRs: e.g., Shrapnel A in Vela SNR \citep{miya01} and those found 
in Cassiopeia A \citep{hugh00}.

While the possibility of Type Ia origin for N49 is intriguing, we caution that the utility of 
our $M_{\rm Si}$/$M_{\rm S}$ estimate to identify the SN type is limited, because it is based 
on small localized ejecta features whereas SN nucleosynthesis model calculations are for the 
integrated ejecta material from the entire SN. Thus, based on the current data, our discussion 
on the origin of N49 is far from conclusive. A more extensive ejecta search and comprehensive 
nucleosynthesis study are required to reveal the true origin of N49. We point out that the 
correct identification of N49's origin is particularly important because of its astrophysical 
implications in the context of the SNR's environment. For instance, if N49 is the remnant of 
a Type Ia SN, the long-standing argument for its physical association with SGR 0526--66 is 
unambiguously ruled out. A Type Ia origin for N49 may also suggest an intriguing case for a 
prompt population Ia SN from a relatively young white dwarf progenitor in analogy to the 
scenario suggested for SNR 0104--72.3 in the Small Magellanic Cloud \citep{lee11}. 

\subsection{\label{subsec:disc_sgr} SGR 0526--66}

Our deep {\it Chandra} observation firmly establishes the previously suggested two-component 
X-ray spectrum for SGR 0526--66. The PL index ($\Gamma$ $\sim$ 2.5) of its X-ray spectrum is 
intermediate between those for other SGRs ($\Gamma$ $\sim$ 1--2) and AXPs ($\Gamma$ $\sim$ 3--4), 
which suggests that SGR 0526--66 may be a transition object between the SGR and AXP states 
\citep{kulk03}. In fact, our estimated PL index ($\Gamma$ $\sim$ 2.5) is identical to that of 
AXP 1E 1048.1--5937, which is considered to be a clear case of a SGR-AXP transition object 
\citep{gavr02}. The implied size of the BB radiation ($R_{\rm BB}$ $\sim$ 5--6 km) is smaller 
than the canonical size of neutron stars. The small $R_{\rm BB}$ suggests the existence of 
restricted hot spots on the surface of the neutron star.

Alternatively, a BB+BB model can equally describe the observed spectrum of 0526--66. The 
two-component BB model was preferred for some AXPs, because the second BB component is physically 
more reasonable than the steep PL component to explain the low flux limit at longer wavelengths 
\citep[e.g., ][]{halp05}. Recently, Nakagawa et al. (2009) showed that the X-ray spectrum of 
the quiescent emission from 
0526--66 can be fitted by a two-component BB model. Our result is consistent with that by Nakagawa 
et al. (2009). The hard BB component indicates a small emission area ($R$ $\sim$ 1 km) with a hot 
temperature of $kT$ $\sim$ 1 keV. The soft BB component indicates an area corresponding to the 
entire surface of the neutron star ($R$ $\sim$ 10 km) with a lower temperature of $kT$ $\sim$ 
0.4 keV. These overall characteristics are reminiscent of the peculiar types of neutron stars 
found at the center of several young SNRs \citep[e.g., ][]{pavl00,park09}, but the estimated BB 
temperatures of 0526--66 are significantly higher than those estimated in others. 

\begin{figure}[]
\figurenum{5}
\centerline{\includegraphics[angle=0,width=\textwidth]{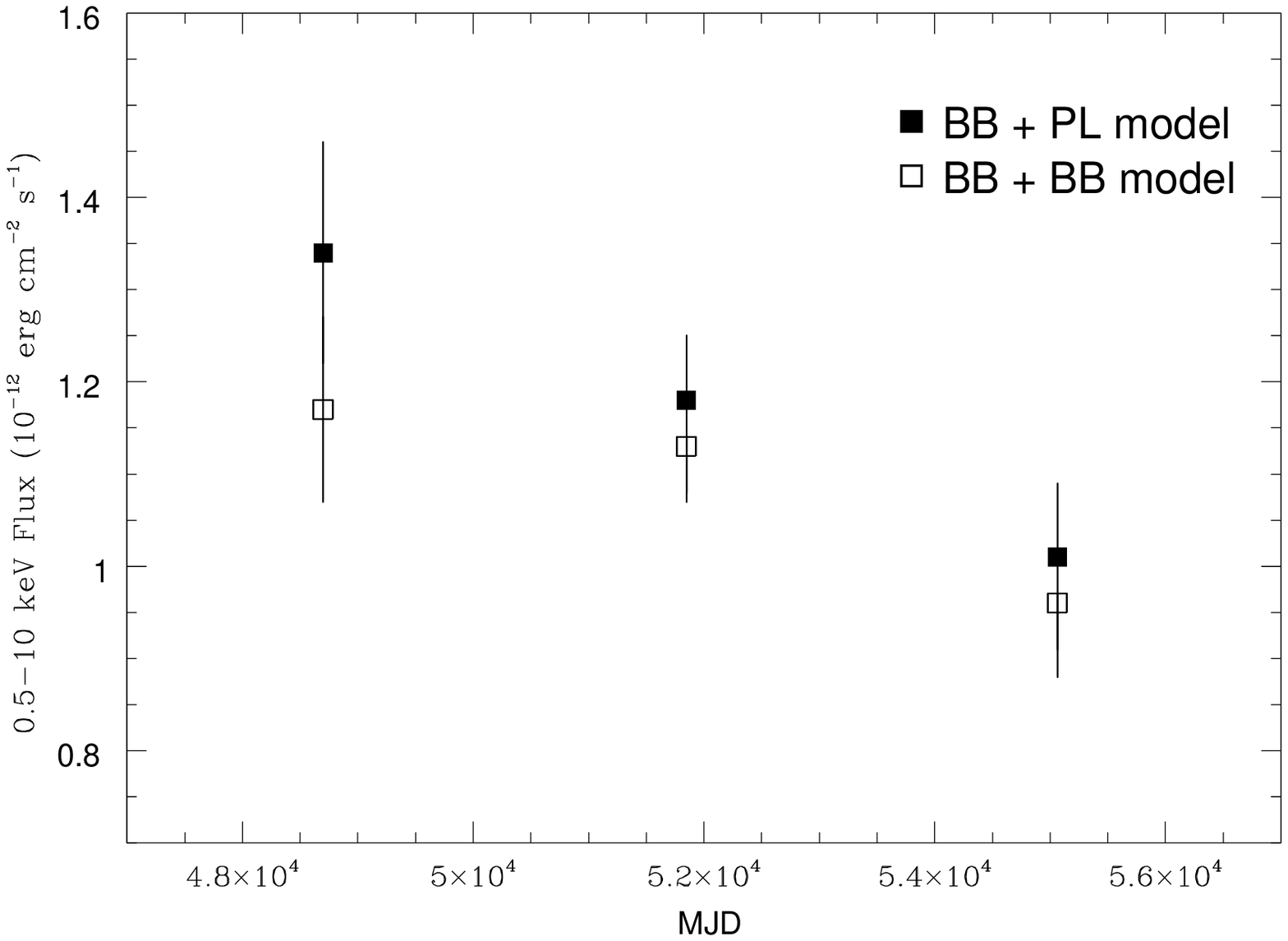}}
\figcaption[]{The long-term X-ray light curve of SGR 0526--66. Fluxes at the first epoch 
have been estimated by the {\it ROSAT} HRI data \citep{roth94}. 1$\sigma$ error bars based 
on the count statistics are shown in the {\it ROSAT} fluxes. The next two fluxes are estimated 
by {\it Chandra} data. Two model fits (BB+PL and BB+BB) are assumed for the flux estimates.
For the {\it Chandra} fluxes, uncertainties are with a 90\% confidence level as estimated
by ``flux error'' command in the XSPEC based on the two-component model fits.
\label{fig:fig5}}
\end{figure}

The overall X-ray flux in the 0.5--10 keV band in 2009 is $\sim$15\% lower than it was in 
2000--2001.  We show a long-term X-ray light curve of SGR 0526--66 in Figure~\ref{fig:fig5}. 
In Figure~\ref{fig:fig5}, we plot the mean flux of 2000 and 2001 data (the middle data point) 
because the observed flux is indistinguishable between the two epochs. We added the fluxes 
estimated by the {\it ROSAT} HRI data in this light curve. For the {\it ROSAT} fluxes, we 
converted the observed HRI count rate \citep[1.51$\pm$0.13 $\times$ 10$^{-2}$ counts s$^{-1}$,
][]{roth94} into the 0.5--10 keV band energy flux using PIMMS. In this calculation, we assumed 
two cases of BB+PL and BB+BB models with the best-fit parameters listed in Tables~\ref{tbl:tab4} 
and \ref{tbl:tab5}, respectively. We also assumed the fractional contributions in the observed 
0.1--2.4 keV band HRI count rate from each of the model components, based on the results 
summarized in Table~\ref{tbl:tab4} and \ref{tbl:tab5}. The calculated {\it ROSAT} HRI fluxes 
are $f_{\rm 0.5-10~keV}$ $\sim$ 1.34 (BB+PL) and $\sim$1.17 (BB+BB) $\times$ 10$^{-12}$ erg 
cm$^{-2}$ s$^{-1}$. While the X-ray flux appears to have decreased by $\sim$30\% since 1992 
based on the BB+PL case, the overall flux decrease is less certain for the BB+BB case. If it is 
real, the suggested X-ray flux change for 0526--66 would not be surprising, because long-term 
variabilities by a factor of up to $\sim$10 in several years have been detected in some AXPs 
\citep[e.g., ][]{bayk96,oost98}. Follow-up {\it Chandra} observations are essential to reveal 
the true nature of the long-term light curve of SGR 0526--66.    

We note that the foreground column to 0526--66 shows a significant discrepancy by a factor of 
$\sim$3 between values estimated by two different spectral modelings (BB+PL vs. BB+BB). The 
column estimated by the BB+BB model fit ($N_{\rm H, LMC}$ $\sim$ 1.7 $\times$ 10$^{21}$ cm$^{-2}$) 
is generally in agreement with those measured for N49 (Table~\ref{tbl:tab2}), and particularly, 
it is fully consistent with the column toward the SGR\_BG region ($N_{\rm H, LMC}$ $\sim$ 1.6 
$\times$ 10$^{21}$ cm$^{-2}$). These consistent columns between N49 and 0526--66 are supportive 
of the long-suggested physical association between them. On the other hand, our BB+PL model fit 
of 0526--66 spectrum shows a substantially larger foreground column for SGR 0526--66 ($N_{\rm 
H,LMC}$ $\sim$ 5.4 $\times$ 10$^{21}$ cm$^{-2}$) than that for SGR\_BG region. If this is the 
case, the large difference in the foreground absorption between SNR N49 and SGR 0526--66 brings 
into question their physical association. If we assumed an average interstellar density of $n_0$ 
$\sim$ 1--2 cm$^{-3}$ (see Section~\ref{subsec:disc_ejecta}) near N49, SGR 0526--66 may be 
$\sim$500--1000 pc beyond N49. Thus, this model-dependent difference in $N_{\rm H, LMC}$ for 
0526--66 deserves full attention for further studies with follow-up observations. An H {\small
I} survey of the LMC shows that N49 and 0526--66 are projected toward the dense boundary 
between two supergiant shells \citep{kim03}. The estimated  H {\small I} column density 
toward this region appears to be $N_{\rm H, LMC}$ $\sim$ 5 $\times$ 10$^{21}$ cm$^{-2}$. 
It is difficult to discriminate our model-dependent $N_{\rm H, LMC}$ measurements toward 
0526--66 based on this relatively large column estimated by H {\small I} data with a poor 
angular resolution ($\sim$1$'$ which is comparable with the angular size of N49). We note 
that, if 0526--66 and N49 are associated, the projected angular separation 
($\sim$22$^{\prime\prime}$) of 0526--66 from the geometric center of N49 requires a 
large kick-velocity (i.e., an average transverse velocity of $v$ $\sim$ 1100 km s$^{-1}$ 
for the SNR age of $\sim$4800 yr). At least, the proper measurement of the foreground 
column toward 0526--66 is directly related with two important astrophysical issues: (1) 
the origin of the quiescent X-ray emission from SGR 0526--66, and (2) the physical 
relationship between SNR N49 and SGR 0526--66. It may also provide a useful piece of 
puzzle to reveal the origin of SNR N49 (thermonuclear vs. core-collapse).
 
\section{\label{sec:sum} SUMMARY}

Using our deep {\it Chandra} observation, we detect metal-rich ejecta features in SNR N49.
These ejecta features include a ``bullet'' that is most likely a stellar fragment travelling 
beyond the southwestern boundary of the SNR. We also find an ejecta feature in the eastern 
part of the SNR, nearly in the opposite side of the SNR to the bullet. Both of these ejecta
features show highly enhanced Si and S abundances by an order of magnitude above the LMC values. 
We do not find compelling evidence for overabundant O and/or Fe in these ejecta features. If 
N49 is a remnant of a core-collapse explosion of a massive progenitor, the Si-rich nature of
ejecta may be considered to be nucleosynthesis products of explosive O-burning or incomplete 
Si-burning from deep inside of the SN. On the other hand, the estimated Si- and S-rich ejecta 
mass ratio appears to favor a Type Ia origin for N49. If N49 was the remnant of a Type Ia SN, 
the suggested physical association between N49 and SGR 0526--66 would be unambiguously ruled 
out. However, we note that our SN ejecta study is limited because we used only some small 
localized ejecta features rather than the integrated ejecta composition from the entire SNR 
to represent the true SN nucleosynthesis. Follow-up studies to reveal the comprehensive ejecta 
composition in N49 are required to unveil the true origin of this SNR.  

Our new {\it Chandra} observation allows us to detect the blast wave forming the swept-up shell 
in the southern boundary with significant photon statistics. Since this part of the shell is not 
affected by spectral complications due to the SNR's interaction with dense clumpy clouds, it 
provides a useful opportunity to reveal the dynamics of the SNR more accurately than previous
works. We estimate the Sedov age of $\tau_{\rm Sed}$ $\sim$ 4800 yr and the explosion energy
$E_0$ $\sim$ 1.8 $\times$ 10$^{51}$ erg for N49.

Our spectral analysis of the quiescent X-ray emission from SGR 0526--66 using the deep exposure 
clearly reveals the presence of a BB emission ($kT_{\rm BB}$ $\sim$ 0.44 keV) in addition to a PL 
component. The implied BB emitting area is relatively small ($R$ $\sim$ 5--6 km) compared to the 
canonical size of neutron stars. The estimated PL photon index ($\Gamma$ $\sim$ 2.5) is identical 
to that of AXP 1E 1048.1--5937, the well-known candidate transition object between AXPs and SGRs. 
Alternatively, the observed X-ray spectrum of 0526--66 can be equally fitted by a two-component
BB model. This model indicates that X-ray emission originates from a small ($R$ $\sim$ 1 km), hot 
($kT$ $\sim$ 1 keV) spot(s) in addition to a cooler ($kT$ $\sim$ 0.4 keV) surface of the neutron
star ($R$ $\sim$ 10 km).

We find marginal evidence for a slow decay in the observed X-ray flux of 0526--66 ($\sim$20--30\%
for the last $\sim$17 yr). Continuous X-ray monitoring of 0526--66 is needed to clarify the nature 
of its long-term light curve. We find a considerable difference in the foreground column by a 
factor of $\sim$3 between two modelings (BB+PL vs. BB+BB) of the X-ray spectrum of 0526--66. 
While such a model-dependent discrepancy has been noticed for several other AXPs, the 0526--66 
case is particularly intriguing because discriminating these competing models may be able to 
provide some critical clues on the nature of 0526--66 and N49: e.g., their physical association, 
the origin of X-ray emission of 0526--66, and the origin of N49.

\acknowledgments

This work has been supported in part by the SAO under the {\it Chandra} grants GO9-0072A
to the University of Texas at Arlington, and GO9-0072X to the Pennsylvania State University. 
J.P.H. has been supported in part by the {\it Chandra} grant GO0-11090X.

\clearpage

\begin{deluxetable}{cccc}
\footnotesize
\tablecaption{{\it Chandra} Observations of N49
\label{tbl:tab1}}
\tablewidth{0pt}
\tablehead{ \colhead{ObsID} & \colhead{Observation Date} & \colhead{Exposure (ks)} & 
\colhead{Instrument} }
\startdata
10123 & 2009-7-18 & 26.8 & ACIS-S3 (1/4 subarray) \\
10806 & 2009-9-19 & 26.5 & ACIS-S3 (1/4 subarray) \\
10807 & 2009-9-16 & 26.0 & ACIS-S3 (1/4 subarray) \\
10808 & 2009-7-31 & 28.7 & ACIS-S3 (1/4 subarray) \\
747 & 2000-1-4 & 39.9 & ACIS-S3 (1/8 subarray) \\
1957 & 2001-8-31 & 48.4 & ACIS-S3 (1/8 subarray) \\
\enddata
\tablecomments{In the {\it Chandra} archive, there are two other ObsIDs (1041 and 2515) that 
detected N49. Those observations are not useful for the purposes of this work because of a large 
off-axis pointing ($\sim$6$\farcs$5 for ObsID 1041) or a short exposure (7 ks for ObsID 2515). 
Thus, we excluded them in this work.
}

\end{deluxetable}

\clearpage

\begin{deluxetable}{cccccc}
\footnotesize
\tablecaption{Summary of Spectral Model Fits to Subregions of N49
\label{tbl:tab2}}
\tablewidth{0pt}
\tablehead{ \colhead{} & \colhead{$N_{\rm H,LMC}$} & \colhead{$kT$} & \colhead{$n_et$}
& \colhead{$EM$\tablenotemark{a}} & \colhead{$\chi^2$/n} \\
\colhead{Region} & \colhead{(10$^{21}$ cm$^{-2}$)} & \colhead{(keV)} & \colhead{(10$^{11}$ 
cm$^{-3}$ s)} & \colhead{(10$^{57}$ cm$^{-3}$)} & \colhead{} }
\startdata
Shell & 0.89$^{+1.31}_{-0.79}$ & 0.57$^{+0.05}_{-0.10}$ & 6.35$^{+9.65}_{-3.90}$ &
6.75$^{+6.78}_{-2.73}$ & 49.1/68 \\
SGR\_BG & 1.58$^{+0.45}_{-0.44}$ & 0.56$\pm$0.03 & 9.65$^{+6.55}_{-4.35}$ & 
15.45$^{+3.66}_{-3.06}$ & 258.9/218 \\
Tail\tablenotemark{b} & 1.01$^{+2.60}_{-1.00}$ & 2.02$^{+1.41}_{-0.24}$ & 0.68$^{+0.58}_{-0.30}$ &
0.50$^{+0.23}_{-0.28}$ & 44.9/44 \\
Tail\tablenotemark{c}  & 1.68$^{+1.82}_{-1.48}$ & 0.71$^{+0.11}_{-0.07}$ & 4.02$^{+3.78}_{-2.07}$ &
1.61$^{+0.40}_{-0.28}$ & 52.4/49 \\
Head & 1.32$^{+0.60}_{-0.57}$ & 1.04$^{+0.06}_{-0.05}$ & $>$20.3 & 2.25$^{+0.19}_{-0.20}$ &
67.0/59 \\
East\tablenotemark{d} & 3.11$^{+0.81}_{-0.88}$ & 1.09$^{+0.05}_{-0.04}$ & $>$67.0 & 
2.68$^{+0.20}_{-0.24}$ & 144.4/133 \\
\enddata
\tablecomments{Uncertainties are with a 90\% confidence level. The 90\% limit is presented
where the best-fit value is unconstrained. The Galactic column $N_{\rm H,Gal}$ 
is fixed at 0.6 $\times$ 10$^{21}$ cm$^{-2}$ \citep{dick90}.}
\tablenotetext{a}{Volume emission measure, $EM$ = $\int n_e n_H dV$, assuming the distance to the 
LMC, $d$ = 50 kpc.}
\tablenotetext{b}{The best-fit parameters are from a single shock model fit (see text).}
\tablenotetext{c}{The best-fit parameters are for the soft component from a two-temperature 
model fit (see text).}
\tablenotetext{d}{The best-fit parameters for the ejecta component. }
\end{deluxetable}

\begin{deluxetable}{ccccccc}
\footnotesize
\tablecaption{Summary of Estimated Metal Abundances in Subregions of N49
\label{tbl:tab3}}
\tablewidth{0pt}
\tablehead{ \colhead{Region} & \colhead{O} & \colhead{Ne} & \colhead{Mg}
& \colhead{Si} & \colhead{S} & \colhead{Fe} }
\startdata
Shell & 0.32$^{+0.49}_{-0.18}$ & 0.27$^{+0.21}_{-0.11}$ & 0.25$\pm$0.10 &
0.36$^{+0.17}_{-0.14}$ & 0.84$^{+0.69}_{-0.55}$ & 0.21$^{+0.12}_{-0.08}$ \\
SGR\_BG & 0.24$^{+0.11}_{-0.09}$ & 0.17$^{+0.07}_{-0.06}$ & 0.15$^{+0.04}_{-0.03}$ &
0.24$^{+0.06}_{-0.06}$ & 0.59$^{+0.26}_{-0.22}$ & 0.13$^{+0.02}_{-0.01}$ \\
Tail & 0.28$^{+0.72}_{-0.14}$ & $<$1.54 (0.58) & 0.37$^{+0.56}_{-0.19}$ &
0.88$^{+1.24}_{-0.38}$ & $<$2.30 (0.49) & 0.28$^{+0.87}_{-0.16}$ \\
Head\tablenotemark{a} & 0.32 & 0.27 & 0.25 & 2.28$^{+0.48}_{-0.39}$ & 3.22$^{+0.90}_{-0.80}$ 
& 0.21 \\
East\tablenotemark{b} & 0.27 & 0.21 & 0.20 & 1.92$^{+0.29}_{-0.26}$ & 1.39$^{+0.44}_{-0.39}$ & 0.17 \\
\enddata
\tablecomments{Abundances are with respect to Solar \citep{ande89}. Uncertainties are with 
a 90\% confidence level. The 90\% limit is presented where the best-fit value is unconstrained. }
\tablenotetext{a}{Abundances for O, Ne, Mg, and Fe are fixed at values measured for the ``Shell'' 
region. }
\tablenotetext{b}{Abundances for O, Ne, Mg, and Fe are fixed at the average values measured for 
the ``Shell'' and ``SGR\_BG'' regions. }
\end{deluxetable}

\begin{deluxetable}{lccc}
\footnotesize
\tablecaption{Summary of PL+BB spectral model fit to SGR 0526--66
\label{tbl:tab4}}
\tablewidth{0pt}
\tablehead{ \colhead{Parameter} & \colhead{BB} & \colhead{PL} & \colhead{Overall}}
\startdata
$kT$ (keV) & 0.44$\pm$0.02 & - & - \\
$\Gamma$ & - & 2.50$^{+0.11}_{-0.12}$ & - \\
$R_{\rm BB,1}$\tablenotemark{a} (km) & 6.0$^{+0.7}_{-0.6}$ & - & - \\
$R_{\rm BB,2}$\tablenotemark{b} (km) & 5.5$\pm$0.6 & - & - \\
$f_{\rm X,1}$\tablenotemark{a} (10$^{-12}$ erg cm$^{-2}$ s$^{-1}$) & 0.40$^{+0.10}_{-0.08}$ &
0.78$\pm$0.13 & 1.18$^{+0.10}_{-0.15}$ \\
$f_{\rm X,2}$\tablenotemark{b} (10$^{-12}$ erg cm$^{-2}$ s$^{-1}$) & 0.33$^{+0.08}_{-0.07}$ &
0.69$\pm$0.11 & 1.01$^{+0.08}_{-0.13}$ \\
$L_{\rm X,1}$\tablenotemark{a} (10$^{35}$ erg s$^{-1}$) & 1.68$^{+0.41}_{-0.35}$ & 3.84$\pm$0.65
& 5.52$^{+0.48}_{-0.70}$ \\
$L_{\rm X,2}$\tablenotemark{b} (10$^{35}$ erg s$^{-1}$) & 1.37$^{+0.32}_{-0.28}$ & 3.36$\pm$0.54
& 4.74$^{+0.38}_{-0.61}$ \\
$N_{\rm H,LMC}$ (10$^{21}$ cm$^{-2}$) & - & - & 5.44$^{+0.58}_{-0.59}$ \\
$N_{\rm H,Gal}$ (10$^{21}$ cm$^{-2}$) & - & - & 0.6 (fixed) \\
$\chi^2$/n & - & - & 577.4/582 \\
\enddata
\tablecomments{Uncertainties are with a 90\% confidence level.}
\tablenotetext{a}{This is based on the {\it Chandra} data taken in 2000 and 2001. $f_{\rm X}$ and 
$L_{\rm X}$ are estimated in the 0.5--10 keV band. The distance to the LMC, $d$ = 50 kpc, is assumed.}
\tablenotetext{b}{This is based on the {\it Chandra} data taken in 2009. $f_{\rm X}$ and $L_{\rm X}$ 
are estimated in the 0.5--10 keV band. The distance to the LMC, $d$ = 50 kpc, is assumed.}
\end{deluxetable}

\begin{deluxetable}{lccc}
\footnotesize
\tablecaption{Summary of BB+BB spectral model fit to SGR 0526--66
\label{tbl:tab5}}
\tablewidth{0pt}
\tablehead{ \colhead{Parameter} & \colhead{BB$_{\rm soft}$} & \colhead{BB$_{\rm hard}$} 
& \colhead{Overall}}
\startdata
$kT$ (keV) & 0.39$\pm$0.01 & 1.01$^{+0.11}_{-0.09}$ & - \\
$R_{\rm BB,1}$\tablenotemark{a} (km) & 9.7$^{+0.6}_{-0.5}$ & 1.0$\pm$0.2 & - \\
$R_{\rm BB,2}$\tablenotemark{b} (km) & 9.0$\pm$0.5 & 0.9$\pm$0.2 & - \\
$f_{\rm X,1}$\tablenotemark{a} (10$^{-12}$ erg cm$^{-2}$ s$^{-1}$) & 0.73$^{+0.09}_{-0.07}$ &
0.40$^{+0.22}_{-0.15}$ & 1.13$^{+0.01}_{-0.05}$ \\
$f_{\rm X,2}$\tablenotemark{b} (10$^{-12}$ erg cm$^{-2}$ s$^{-1}$) & 0.62$^{+0.34}_{-0.23}$ &
0.35$^{+0.19}_{-0.13}$ & 0.96$^{+0.01}_{-0.05}$ \\
$L_{\rm X,1}$\tablenotemark{a} (10$^{35}$ erg s$^{-1}$) & 0.90$^{+0.11}_{-0.09}$ & 
0.42$^{+0.23}_{-0.16}$ & 3.96$^{+0.05}_{-0.18}$ \\
$L_{\rm X,2}$\tablenotemark{b} (10$^{35}$ erg s$^{-1}$) & 0.77$^{+0.43}_{-0.28}$ & 
0.36$^{+0.20 }_{-0.13}$ & 3.39$^{+0.04}_{-0.16}$ \\
$N_{\rm H,LMC}$ (10$^{21}$ cm$^{-2}$) & - & - & 1.70$^{+0.25}_{-0.23}$ \\
$N_{\rm H,Gal}$ (10$^{21}$ cm$^{-2}$) & - & - & 0.6 (fixed) \\
$\chi^2$/n & - & - & 616.4/582 \\
\enddata
\tablecomments{Uncertainties are with a 90\% confidence level.}
\tablenotetext{a}{This is based on the {\it Chandra} data taken in 2000 and 2001. $f_{\rm X}$ and 
$L_{\rm X}$ are estimated in the 0.5--10 keV band. The distance to the LMC, $d$ = 50 kpc, is assumed.}
\tablenotetext{b}{This is based on the {\it Chandra} data taken in 2009. $f_{\rm X}$ and $L_{\rm X}$ 
are estimated in the 0.5--10 keV band. The distance to the LMC, $d$ = 50 kpc, is assumed.}
\end{deluxetable}

\end{document}